\documentclass[%
 reprint,
 amsmath,amssymb,
 aps,
 prx,
showkeys]{revtex4-2}

\usepackage{graphicx}
\usepackage{dcolumn}% Align table columns on decimal point
\usepackage{bm}% bold math
\usepackage{amsfonts, amsmath, amssymb}
\newcommand{\ket}[1]{|#1\rangle}

\newcommand{\bra}[1]{\langle #1|}

\newcommand{\braket}[2]{\langle #1 | #2 \rangle}

\newcommand{\vacp}{\ket{0}\bra{0}}

\newcommand{\pd}[2]{\frac{\partial #1}{\partial #2}}

\begin{document}

\title{Towards the Quantum Limits of Phase Retrieval}

\author{Jacob Trzaska}
\affiliation{James C. Wyant College of Optical Sciences, University of Arizona, Tucson, Arizona, USA}

\author{Amit Ashok}
\affiliation{James C. Wyant College of Optical Sciences, University of Arizona, Tucson, Arizona, USA}
\affiliation{Department of Electrical and Computer Engineering, University of Arizona, Tucson, Arizona, USA}
\date{\today}

\begin{abstract}
We consider the problem of determining the spatial phase profile of a single-mode electromagnetic field. Our attention is on input states that are a statistical mixture of displaced and squeezed number states, a superset of Gaussian states. In particular, we derive the quantum Fisher information matrix (QFIM) for estimating the expansion coefficients of the wavefront in an orthonormal basis, finding that it is diagonal. Moreover, we show that a measurement saturating the QFIM always exists, and point to an adaptive strategy capable of implementing it. We then construct the optimal measurements for three particular states: mixtures of photon number, coherent, and single-mode squeezed vacuum states. Sensitivity of the measurements to nuisance parameters is explored.
\end{abstract}

\keywords{Optics, Quantum Information, Phase Retrieval, Wavefront Sensing} %Use showkeys class option if keyword display desired

\maketitle

%\tableofcontents

\section{\label{sec:intro}INTRODUCTION}
Phase retrieval (or estimation) is the problem of determining the spatial phase profile of an electromagnetic field from coded irradiance measurements. Problems of this variety are exceedingly common; examples include astronomical wavefront sensing \cite{guyon2018extreme}, optical testing \cite{briers1999optical}, microscopy \cite{booth2014adaptive}, x-ray crystallography \cite{millane1990phase}, and gravitational wave astronomy \cite{caves_quantum-mechanical_nodate}. However, quantifying the fundamental performance limits to phase retrieval is complicated by disparate sources employed across various measurement systems, e.g., lasers in optical testing and thermal light in astronomy. Indeed, the quantum mechanical behavior of lasers and thermal sources is remarkably different. One would expect that the optimal phase-retrieving measurement is highly dependent on the nature of the source. By considering the of quantum nature of light, our goal is to present a unified treatment of phase retrieval methods informed by a rigorous quantum information theoretical analysis.

Several prior works \cite{pinel_ultimate_2012, jarzyna_quantum_2012, demkowicz-dobrzanski_elusive_2012, humphreys_quantum_2013, sparaciari_bounds_2015, nichols_multiparameter_2018} have applied the tools of quantum metrology to the phase measurement problem in optical interferometry. For example, the problem of estimating a single phase difference across a two arm interferometer has been analyzed for the case of Gaussian states of light. General multiple-parameter phase estimation remained an outstanding challenge due to possible non-commutativity between the unitary operators describing the phase shifts, and possible non-commutativity between the optimal measurements. Extensions to multiple phase estimation \cite{haffert2023} have been recently considered by the astronomical community for estimating the coefficients of an optical wavefront in an orthonormal basis. However, these results have limited scope due to restrictive assumptions about the source, and offer no concrete procedure for constructing the optimal positive operator-valued measure (POVM). More recent work \cite{villegas_optimal_2024} focusing on arbitrary pure states has investigated the structure of the quantum Fisher information matrix (QFIM) for estimating multiple parameters encoded in a spatial phase profile, highlighting work by Pezze \cite{pezze_optimal_2017} for constructing the QFIM-saturating measurement.

In this context, our contributions are twofold. First, we generalize these prior results for phase estimation to statistical mixtures of displaced and squeezed number states. Note that this class includes the set of Gaussian states often used in continuous variable quantum information \cite{weedbrook_gaussian_2012, braunstein_quantum_2005}. In particular, we calculate the QFIM for sensing the wavefront expansion coefficients as expressed in an orthonormal basis. Via a Jacobian transformation, one can then calculate the QFIM for any reparametrization of the mode coefficients \cite{liu_quantum_2020}. Second, we prove that a POVM capable of saturating the QFIM \textit{always} exists. However, the POVM contains an explicit dependence on the parameter values, thereby presenting a measurement design challenge in practical settings. Strategies to circumvent this parameter-dependence are also discussed, followed by an analysis characterizing the sensitivity of the measurement to perturbations in nuisance parameters. 

\section{\label{sec:theory}THEORY}
\subsection{Notation}
Let $\mathcal{P}\subset\mathbb{R}^2$ denote the clear aperture/pupil of an optical system and $f, g : \mathcal{P}\rightarrow \mathbb{C}$ be two complex-valued function defined over the pupil. We define an inner product between $f$ and $g$ as \begin{equation}
    \label{eq:ip}
    (f, g) = \int_{\mathcal{P}}{f^*(\mathbf{x})g(\mathbf{x})\gamma(\mathbf{x)}d\mathbf{x}},
\end{equation} where the integration is over $\mathcal{P}$, $*$ denotes complex conjugation, and $\gamma(\mathbf{x})$, is a real-valued weighting function. With equation (\ref{eq:ip}) we can construct an orthonormal basis for the vector space of functions defined over $\mathcal{P}$. 

Our analysis will consider only a single spatiotemporal mode of the electromagnetic field. Such a mode admits the position-space representation  \begin{equation}
    \omega_0(\mathbf{x}) = u(\mathbf{x})e^{iW(\mathbf{x})},
\end{equation} where \begin{equation}
    W(\mathbf{x}) = \sum_{n=0}^{\infty}\alpha_n\varphi_n(\mathbf{x}),\,\,\,\alpha_n\in\mathbb{R}
\end{equation} is the phase of the field and $u(\mathbf{x})$ is a normalized amplitude: \begin{equation}
    \int_{\mathcal{P}}{|u(\mathbf{x})|^2dx} = 1.
\end{equation} The set of basis functions $\{\varphi_{n}\}_{n=0}^{\infty}$ are defined over $\mathcal{P}$, and are chosen to be orthonormal with respect to the integral in equation (\ref{eq:ip}), with the substitution $\gamma = |u|^2$. The mode $\varphi_0$, taken here as constant, represents a global phase and will be ignored in subsequent analysis. 

Note that the amplitude $u$ is assumed to be known; our focus is exclusively on $W$. Also, for apodized pupils the apodization function can be simply incorporated into the mode definition, rendering the clear aperture assumption unchanged. 

We now define the modes \begin{equation}
    \label{eq:modes}
    \omega_n = \begin{cases}
        \omega_0, & n = 0 \\
        \pd{\omega_0}{\alpha_n}, & n > 0.
    \end{cases}
\end{equation} This collection forms an orthonormal set over $\mathcal{P}$, and is a natural basis for the ensuing quantum information analysis, as only states and their derivatives span the space that will be relevant to the phase estimation problem. Each of these modes is assigned a complex Hilbert space with creation and annihilation operators $\hat{a}^\dagger_n$ and $\hat{a}_n$, respectively. These operators obey the usual algebra \begin{subequations}
    \label{eq:modal-commutators}
    \begin{align}
        \left[\hat{a}_i, \hat{a}_j\right] &= 0, \\
        \left[\hat{a}_i, \hat{a}^\dagger_j\right] &= \delta_{ij},
    \end{align}
\end{subequations} where $[\cdot,\cdot]$ is the commutator product. A state consisting of $n$ photons in mode $\omega_k$ will be denoted by $\ket{n}_k$.

Lastly, we will use $\partial_l$ to denote the derivative with respect to the $l^{\text{th}}$ modal coefficient, i.e., \begin{equation}
    \partial_l = \frac{\partial}{\partial\alpha_l}.
\end{equation}

\subsection{Wavefront Sensing with Statistical Mixtures of Displaced and Squeezed Number States}
Let \begin{align}
    \mu &= |\mu|e^{i\kappa}\\
    \xi &= re^{i\theta},
\end{align} where $\theta, \kappa\in[0,2\pi)$ and $r, |\mu| \geq 0$. We define a mixed displaced and squeezed number (DSN) state as one admitting the following density operator \begin{equation}
    \label{eq:mixed-dsn}
    \hat{\rho} = \hat{D}(\mu)\hat{S}(\xi)\left(\sum_{n=0}^{\infty}\epsilon_n\ket{n}_0\bra{n}_0\right)\hat{S}^{\dagger}(\xi)\hat{D}^{\dagger}(\mu),
\end{equation} where the $\{\epsilon_n\}$ are real positive constants that sum to unity, and $\hat{S}(\xi)$ and $\hat{D}(\mu)$ denote the squeezing and displacement operators for the $n=0$ mode, respectively (see Appendix \ref{app:useful} for details). The relevance of these states may be made clear with a suitable choice of the $\epsilon_n$. Indeed, upon choosing \begin{equation}
    \epsilon_n = \frac{\langle n\rangle^n}{(\langle n\rangle+1)^{n+1}},
\end{equation} where $\langle n\rangle$ is real and positive, one recovers the class of Gaussian states.

Mixed DSN states are particularly simple to represent. Notice that their eigenstates are those of the form \begin{equation}
    \label{eq:dsn-state}
    \ket{n}'_0 = \hat{D}(\mu)\hat{S}(\xi)\ket{n}_0.
\end{equation} Because $\hat{D}(\mu)\hat{S}(\xi)$ is a unitary operator, the states $\ket{n}'_0$ are orthonormal, and form a basis in which the density matrix is diagonal. Therefore, we can restrict our attention to just these states and their derivatives, as they form the subspace relevant to calculation of the QFIM. Furthermore, the derivatives are readily obtained from that of the squeezed vacuum and displacement operators (Appendix \ref{app:useful}).

We can express the QFIM, $Q$, for sensing the coefficients $\{\alpha_n\}_{n=1}^{\infty}$, in the general form \begin{equation}
    \label{eq:qfi-sum}
    Q_{ij} = 2\sum_{\epsilon_l+\epsilon_k\neq0}\bra{l}'_0\hat{\rho}\ket{k}'_0\frac{\ket{k}'_0\bra{l}'_0}{\epsilon_l+\epsilon_k}.
\end{equation} Interested readers are referred to Appendix \ref{app:dsnQFI} for a detailed evaluation of this sum; we shall quote only the final result: \begin{equation}
    \label{eq:qfi}
    Q_{ij} = 4\left[|\mu|^2+\sinh^2(r) + \langle n\rangle\cosh(2r)\right]\delta_{ij},
\end{equation} Two features of this matrix are particularly noteworthy: (i) the expression in brackets is simply the mean number of photons and (b) only the magnitudes of $\mu$ and $\xi$ determine the quantum limit. Intuitively, this is expected, because only differences in phase are experimentally observable, and the reference phase may also be chosen as zero. Nevertheless, knowledge of the phase does play a crucial role in the design of experiments, a point to which we shall return in Section \ref{sec:measurement-construction}. Finally, note that this result still holds when the order of $\hat{D}(\mu)$ and $\hat{S}(\xi)$ is swapped, up to a transformation of the displacement parameter $\mu$ (Appendix \ref{app:sdn-qfi}).

Having obtained the quantum limit to modal phase estimation, a reasonable next question is \textit{''Is the limit achievable?"}. A necessary and sufficient condition for saturating the QFIM \cite{ ragy_compatibility_2016, demkowicz-dobrzanski_multi-parameter_2020} is weak commutativity, defined as \begin{equation}
    \label{eq:attcond}
    \text{Tr}\left(\hat{\rho}\left[\hat{\mathfrak{L}}_i, \hat{\mathfrak{L}}_j\right]\right) = 0,
\end{equation} where $\hat{\mathfrak{L}}_i$ is the symmetric logarithmic derivative with respect to the $i$-th parameter. We find that for a mixed DSN state this relation is always satisfied (Appendix \ref{app:dsnQFI}). It follows that the quantum limit for modal phase estimation is \textit{always} saturable, although for mixed states the saturation is generally asymptotic, requiring infinite copies of the state. For pure states, however, this result implies saturation of the QFIM at the single copy level.

\subsection{Phase Estimation with Orthogonal Polarization States}
Our results are simply extended to states with polarization-dependent phases. Let us consider the more general electric field mode \begin{equation}
    \label{eq:polarization-mode}
    \vec{\omega}_0(\mathbf{x}) = u_0(\mathbf{x})\vec{p}_0 e^{iW_0(\mathbf{x})} + u_1(\mathbf{x})\vec{p}_1 e^{iW_1(\mathbf{x})},
\end{equation} where $u_0$ and $u_1$ are real, positive and normalized: \begin{equation}
    \int_{\mathcal{P}}\left[|u_0(\mathbf{x})|^2+ |u_1(\mathbf{x})|^2\right]d\mathbf{x} = 1.
\end{equation} Functions $W_0$ and $W_1$ are also real, and $\vec{p}_0, \vec{p}_1$ are a pair of orthogonal polarization vectors: $\vec{p}_k^{\,*}\cdot\vec{p}_l=\delta_{kl}$. As before, each $W$ may be expanded in an orthonormal basis as \begin{equation}
    \label{eq:polarization-expansion}
    W_n(\mathbf{x}) = \sum_{k=0}^{\infty}\alpha_{n, k}\varphi_{n, k}(\mathbf{x}),\,\,\,n=0,1,
\end{equation} where the functions $\{\varphi_{n,k}\}$ obey the orthogonality relation \begin{equation}
    \int_{\mathcal{P}}|u_n(\mathbf{x})|^2\varphi_{n, k}(\mathbf{x})\varphi_{n, l}(\mathbf{x})d\mathbf{x} = \delta_{kl}.
\end{equation}

We again take $\varphi_{0,0}$ and $\varphi_{1,0}$ as constants, but now choose to re-express the first terms of (\ref{eq:polarization-expansion}) as \begin{equation}
    \alpha_{n,0}\varphi_{n, 0} = c_0 +(-1)^n c_1 \varphi_{n,0}^2/\sqrt{2},
\end{equation} where $c_0$ and $c_1$ are real-valued constants. This swaps the polarization-local pistons for an unobservable global phase, $c_0$, and a mode orthogonal to it, built from the terms including $c_1$.

One now notices that the incident mode (equation \ref{eq:polarization-mode}) and it's derivatives are all mutually orthonormal -- formally the integral of the scalar product is a Kronecker symbol -- implying that equations (\ref{eq:modal-commutators}) still hold following quantization. These two relations are all that are necessary to derive the QFIM and to verify that it is achievable (Appendix \ref{app:dsnQFI}). Hence equation (\ref{eq:qfi}) is also the QFIM for the constants $\{c_1\}\cup\{\alpha_{0,k}\}_{k=1}^{\infty}\cup\{\alpha_{1,k}\}_{k=1}^{\infty}$, and the bound is always attainable.

\subsection{Achieving the Quantum Limit}
\label{sec:achieve}
As noted earlier, equation (\ref{eq:attcond}) guarantees the existence of a measurement saturating the quantum bound, but offers no insights towards their construction. Indeed, to our knowledge, no general method exists when the state is mixed. However, for pure states the problem of finding an optimal measurement has been well characterized \cite{pezze_optimal_2017, villegas_optimal_2024}. In particular, \cite{pezze_optimal_2017} provides a detailed procedure for building the optimal POVM.

However, the optimal POVM remains elusive due to it's explicit dependence on the true parameter values ($\mathbf{\Theta}$). This is a common problem in the design of optimal experiments \cite{cochran-fi-utility}; adaptive estimation strategies have been developed to avoid it altogether. One approach is the adaptive quantum state estimation (ASQE) algorithm, first proposed by Nagaoka \cite{nagaoka_quantum_2004, okamoto_experimental_2012}, and proven to be asymptotically optimal by Fujiwara under mild regularity conditions \cite{fujiwara_strong_2011}. ASQE has seen recent theoretical use for estimating the separation between two incoherent point sources \cite{kimizu_adaptive_2024}, and even for quantum phase estimation \cite{rodriguez-garcia_adaptive_2023}. The algorithm has also been demonstrated experimentally using photon polarization as a case study \cite{okamoto_experimental_2012}.

In brief, the ASQE algorithm employs an iterative and adaptive sequence of measurements and maximum likelihood estimates to converge to the true parameters. One begins by selecting arbitrary parameters, $\boldsymbol{\theta}_0$, and constructing the corresponding optimal measurement $M_0(\boldsymbol{\theta}_0)$; the associated probability density is denoted by $\text{Tr}[\hat{\rho}(\boldsymbol{\theta})M_0(\boldsymbol{\theta}_0)]$. A measurement is taken and the log-likelihood \begin{equation}
    \label{eq:asqe-ll0}
    l_0(\boldsymbol{\theta}) = \log \text{Tr}[\hat{\rho}(\boldsymbol{\theta})M_0(\boldsymbol{\theta}_0)]
\end{equation} is optimized, yielding an updated parameter estimate \begin{equation}
    \boldsymbol{\theta}_1 = \underset{\boldsymbol{\theta}}{\text{argmax}}\left[\,l_0(\boldsymbol{\theta})\,\right].
\end{equation} Subsequently, a new measurement is then deployed, now using the POVM optimal for $\boldsymbol{\theta}_1$: $M_1(\boldsymbol{\theta}_1)$. A new estimate is then derived from \begin{align}
    l_1(\boldsymbol{\theta}) &= \log \text{Tr}[\hat{\rho}(\boldsymbol{\theta})M_1(\boldsymbol{\theta}_1)] + l_0(\boldsymbol{\theta})\\
   \boldsymbol{\theta}_2 &= \underset{\boldsymbol{\theta}}{\text{argmax}}\left[\,l_1(\boldsymbol{\theta})\,\right].
\end{align} This process continues indefinitely according to the sequence \begin{align}
    l_{i}(\boldsymbol{\theta}) & = \log \text{Tr}[\hat{\rho}(\boldsymbol{\theta})M_i(\boldsymbol{\theta}_i)] +l_{i-1}(\boldsymbol{\theta})\\
    \boldsymbol{\theta_{i+1}} &= \underset{\boldsymbol{\theta}}{\text{argmax}}\left[\,l_i(\boldsymbol{\theta})\,\right],
\end{align} where $M_i(\boldsymbol{\theta}_i)$ denotes the optimal POVM given the parameter $\boldsymbol{\theta_i}$. Given equation (\ref{eq:attcond}), the sequence \{$M_i\}_i$ should converge to a measurement saturating the quantum Cramer-Rao bound, and $\boldsymbol{\theta}_i\rightarrow\boldsymbol{\Theta}$, at least for small phases. Phase wrapping will otherwise introduce ambiguity in the mode profile, complicating questions of convergence.

Hardware implementation of AQSE (or any adaptive estimator) clearly requires adapting the measurement on each iteration. Recent work  \cite{ozer_reconfigurable_2022} has already demonstrated such a capability using spatial light modulators for demultiplexing multiple spatial modes; the same implementation can in principle be applied to wavefront sensing, insofar that the field approximately resides in a single-photon state. Intense laser light, and other strong sources for which the probability of higher occupation numbers is large will require further optical processing to implement the quantum-optimal POVM. A future communication will further investigate the application of AQSE to phase retrieval -- both in simulation and in experiment.

\section{Analysis of Optimal Measurements for Various Optical Sources}
\label{sec:measurement-construction}
Pure states account for a small fraction of the states encapsulated by equation (\ref{eq:mixed-dsn}), but are sufficient for describing a number of ubiquitous optical sources. Spatial-spectral filtered starlight in astronomical settings, continuous wave laser light in optical testing, and squeezed light in gravitational wave sensing are three examples where phase retrieval (referred to as wavefront sensing in astronomy) is of great practical significance. Results from \cite{pezze_optimal_2017} allow us to readily evaluate the optimal measurement for each of these states, and can be adapted to construct the POVM for any mixed state that is diagonal in the basis of photon number states.

\subsection{Mixtures of Photon Number States}
\label{sec:weak-thermal}
Set $\mu=\xi=0$ in equation (\ref{eq:mixed-dsn}) and consider the state \begin{equation}
    \label{eq:thermal-state}
    \hat{\rho} = \sum_{n=0}^{\infty}\epsilon_n\ket{n}_0\bra{n}_0.
\end{equation} This model is equivalent to specifying a probability mass function \begin{equation}
    P(n) = \epsilon_n
\end{equation} for measuring $n$ photons in the $\omega_0$ mode. Specific cases of this model have appeared extensively in the optical imaging literature (e.g. \cite{tsang_quantum_2016, haffert2023}), usually considering the case for which the single photon state has unit probability. Given our prior analysis, we may immediately write the QFIM simply as \begin{equation}
    Q_{ij} = 4\langle n\rangle\delta_{ij}
\end{equation}

The structure of equation (\ref{eq:thermal-state}) also permits a straightforward calculation of the optimal measurement -- a fortunate result as, again, we know of no general procedure for mixed states. We leave the details to Appendix \ref{app:thermal-povm} and only state the POVM: \begin{align}
    \Pi &= \{\vacp\} \cup\left\{\bigcup_{n=1}^{\infty} \left\{\ket{\Upsilon_{nk}}\bra{\Upsilon_{nk}}\right\}_{k=0}^\infty \right\} \nonumber\\
    &\cup 
    \left\{\mathbb{I} - \vacp - \sum_{n=1}\sum_{k=0}^{\infty}\ket{\Upsilon_{nk}}\bra{\Upsilon_{nk}}\right\},
\end{align} where the kets $\{\Upsilon_{nk}\}_{k=0}^{\infty}$ are defined as \begin{equation}
    \ket{\Upsilon_{nk}} = U_{0k}\ket{n}_0 + \sum_{l=1}^{\infty}U_{lk}\partial_l\ket{n}_0,\,\,U_{0k}\neq0.
\end{equation} The constants $\{U_{lk}\}_{l=0}^{\infty}$ are real, and chosen so that $\{\ket{\Upsilon_{nk}}\}_{k=0}^{\infty}$'s form an orthonormal basis over the space spanned by $\{\ket{n}_0, \partial_1\ket{n}_0, \partial_2\ket{n}_0, \dots\}$.

A useful feature of this measurement is its insensitivity to the mean photon number. Only the modal structure is required for construction. This is in contrast to optimal POVMs for squeezed and coherent states, which are analyzed hereafter.

\subsection{Coherent States}
Coherent states are often used to describe the emission from single-mode lasers, such as the continuous wave lasers prevalent in optical metrology. They are a special case of equation (\ref{eq:mixed-dsn}) for $\xi=0$ and \begin{equation}
    \epsilon_n = \begin{cases}
        1, & n=0 \\
        0, & n>0
    \end{cases}.
\end{equation} From equation (\ref{eq:qfi}) we can immediately write the QFIM as \begin{equation}
    \label{eq:coh-lim}
    Q_{ij} = 4|\mu|^2\delta_{ij}.
\end{equation} 

The scaling with $|\mu|^2$ implies that, with a sufficient number of photons, phase can be characterized to an arbitrary level of precision. However, this is only true with perfect knowledge of $\mu$. If the optimal measurements are instead designed for a parameter $\mu'\in\mathbb{C}$, the classical Fisher information, $F$, of this detuned measurement is given by (Appendix \ref{app:cfim-coherent-nuisance}) \begin{equation}
    F_{ij} = 4|\mu|^2e^{-|\mu-\mu'|^2}\delta_{ij}.
\end{equation}

\begin{figure*}
    \centering
    \includegraphics[scale=0.42]{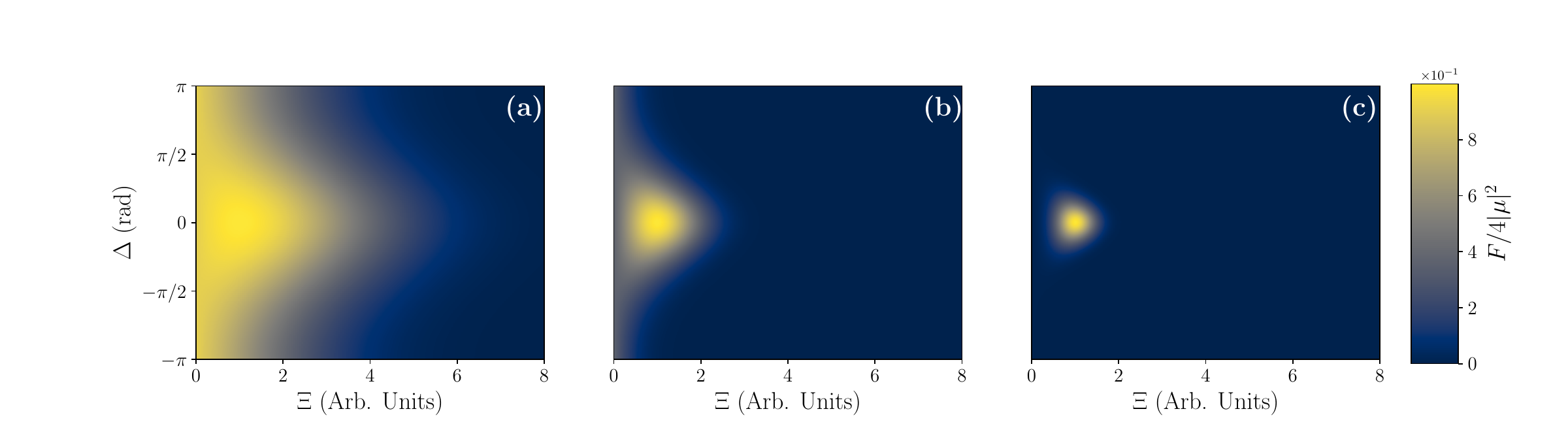}
    \caption{Classical Fisher information for a detuned optimal measurement. The true input states are characterized by mean photon numbers $|\mu|^2$: (a) 0.1, (b) 1 and (c) 10.}
    \label{fig:detuned-coherent-measurement}
\end{figure*}

Note that the loss of sensitivity due to imprecise knowledge of $\mu$ can be quite significant. Let \begin{equation}
    \mu'/\mu = \Xi e^{i\Delta},
\end{equation} describing the measurement detuning. Figure \ref{fig:detuned-coherent-measurement} shows how measurement sensitivity degrades for states with mean photon numbers: (a) 0.1, (b) 1, and (c) 10. At $|\mu|^2=0.1$, detuning the phase by $\pi/4$ leads to nearly twenty percent reduction in achievable sensitivity. As another example, at $|\mu|^2=10$ photons the measurement is rendered effectively insensitive under the same conditions. To put this loss of sensitivity in context, note that a 1 mW HeNe laser (633 nm) delivers about $3\times10^{9}$ ph/$\mu$s, on average. Therefore, ultra-precise beam characterization and exceedingly accurate power stabilization are necessary to build any device capable of working at the quantum limit, as small fluctuations in output power may result in a near-total loss of performance. This suggests that the benefit of operating at the quantum limit may be practically restricted to phase estimation with very weak sources where $|\mu|^2 << 1$, such as very long-standoff imaging and sensing with coherent sources.

\subsection{Single-Mode Squeezed Vacuum}
\label{sec:squeezed}
For our final example, we consider sensing on a single-mode squeezed vacuum state, with squeezing parameter $\xi$. These states correspond to equation (\ref{eq:mixed-dsn}) with $\mu=0$ and \begin{equation}
    \epsilon_n = \begin{cases}
        1, & n=0 \\
        0, & n>0
    \end{cases}.
\end{equation} From equation (\ref{eq:qfi}) we can write the QFIM as \begin{equation}
    \label{eq:squeezed-lim}
    Q_{ij} = 4\sinh^2|\xi|\,\delta_{ij}.
\end{equation} 

\begin{figure*}
    \centering
    \includegraphics[scale=0.44]{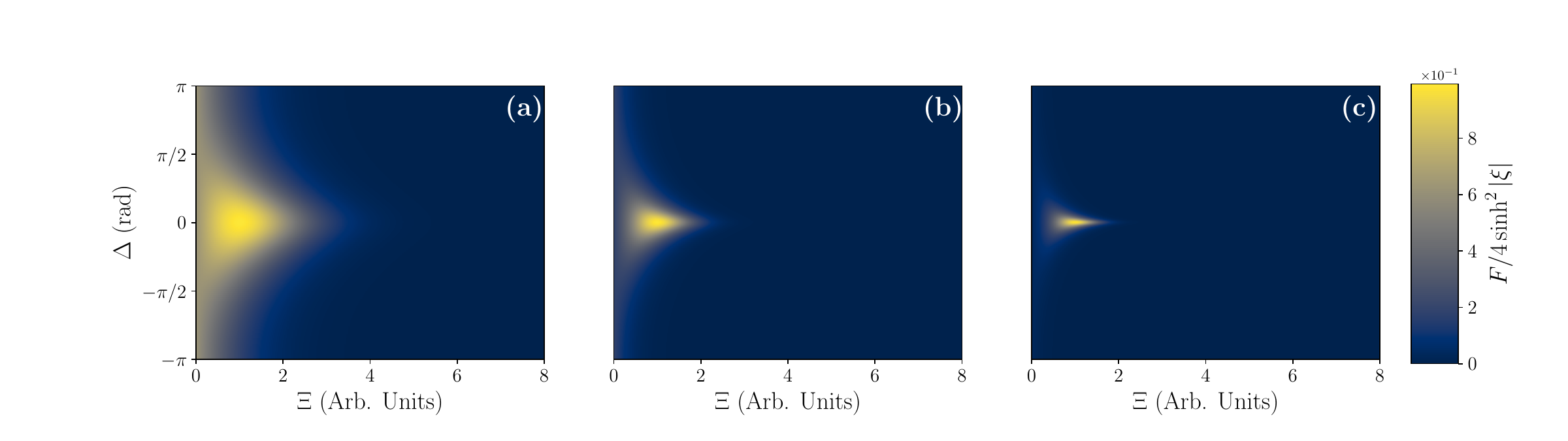}
    \caption{Classical Fisher information for a optimal detuned measurement. The true input states are squeezing at a level (a) 5 dB, (b) 10 dB, and (c) 15 dB.}
    \label{fig:detuned-squeezing-measurement}
\end{figure*}

As with coherent states, the wavefront may be characterized to arbitrary precision by simply increasing the mean photon number -- $\sinh^2|\xi|$ -- and performing the optimal measurement. However, there is again a high cost for improperly characterizing the probe state. Using measurements that are optimal for a squeezing parameter $\xi'$ defined by \begin{equation}
    \xi'/\xi = \Xi e^{i\Delta}
\end{equation} leads to rapid information loss as the detuning is increased (Figure \ref{fig:detuned-squeezing-measurement}). Phase matching is particularly important, a fact that is expected upon considering the overlap of two slightly dephased, and highly squeezed, Wigner functions. Attaining any practical benefit from squeezing will thus require exceptional state characterization prior to modal phase measurements.

\section{Summary and Future Work}
\label{sec:summary}
We have extended existing results in the literature on the precision limits of phase retrieval by determining the quantum Fisher information matrix for statistical mixtures of displaced and squeezed number states -- a superset of the Gaussian states. We have proved that this bound is always achievable, asymptotically in case of a mixed state, and at the single-copy level for pure states. Filtered starlight and continuous-wave laser light, relevant to astronomy and optical testing, respectively, are two examples of common pure states for which the adaptive methods discussed here may be applicable. 

Optimal POVMs for several special states were also analyzed. In general, the optimal measurement depends on the true values of the parameters, motivating an adaptive estimation strategy. In particular, we proposed employing the AQSE algorithm, which produces an estimator whose variance asymptotically saturates the quantum Cramer-Rao bound. Finally, a CFI analysis investigating the role of nuisance parameters for squeezed and coherent states showed that optimally hinges on exceedingly accurate, \textit{a priori} knowledge of the non-estimated parameters, that is the displacement parameter for a coherent state and the squeezing parameter for a single-mode squeezed vacuum state. Fortunately, the quantum-optimal measurements for these parameters are known and may be performed prior to phase estimation.

Further work will relax assumptions about the state of the electromagnetic field. This work specifically assumed a known field amplitude, when in practice it must be estimated alongside the phase. Our assumption of a single-mode field can also be restrictive. An extension to multi-mode fields should be straightforward.

\appendix
\section{Useful Operator Relations}
\label{app:useful}
\subsection{Displacement and Squeezing Operators}
We define our sign convention for the squeezing operator and quote some useful identities for later \cite{glauber_coherent_1963, stoler_equivalence_1970, caves_quantum-mechanical_nodate, loudon2000quantum}. Let \begin{align}
    \hat{D}(\mu) &= \exp\left(\mu\hat{a}^{\dagger}_0-\mu^*\hat{a}_0\right), \\
    \hat{S}(\xi) &= \exp\left\{\frac{1}{2}\left[\xi^*\hat{a}^2_0-\xi(\hat{a}^{\dagger}_0)^2\right]\right\}
\end{align} represent the \textit{displacement} and \textit{squeezing} operators, respectively. The parameters $\mu$ and $\xi$ will be referred to as the displacement and squeezing parameters. 

Displacement operators transform the annihilation operator as \begin{align}
    \label{eq:bog-D_normal}
    \hat{D}^{\dagger}(\mu)\hat{a}_0\hat{D}(\mu) &= \hat{a}_0 + \mu \\
    \label{eq:bog-D_adjoint}
    \hat{D}(\mu)\hat{a}_0\hat{D}^{\dagger}(\mu) &= \hat{a}_0 - \mu.
\end{align} The action of the displacement operator on the vacuum creates a coherent state $\ket{\mu, 0}_0$. Squeezing operators transform the annihilation operator as \begin{align}
    \label{eq:bog-S_normal}
    S^{\dagger}(\xi)\hat{a}_0\hat{S}(\xi) &= \hat{a}_0\cosh(r) - \hat{a}^{\dagger}_0e^{i\theta}\sinh(r) \\
    \label{eq:bog-S_adjoint}
    S(\xi)\hat{a}_0\hat{S}^{\dagger}(\xi) &= \hat{a}_0\cosh(r) + \hat{a}_0^{\dagger}e^{i\theta}\sinh(r).
\end{align} The action of the squeezing operator on the vacuum results in a single-mode squeezed vacuum state $\ket{0, \xi}_0$. Transformations on the creation operator may be found by taking the adjoint of the above relations.

\subsection{Derivative of the Creation Operator, $\hat{a}^{\dagger}_0$, with respect to Modal Phase Coefficients}
\label{app:derivative_creation}
Suppose that we have two orthonormal modal bases $\{\chi_n\}_{n=0}^\infty$ and $\{\chi_n'\}_{n=0}^\infty$. Each set can be associated with a set of creation operators $\{\hat{a}^{\dagger}_n\}_{n=0}^{\infty}$ and $\{\hat{b}^{\dagger}_n\}_{n=0}^{\infty}$, respectively. Converting between mode bases is a matter of appropriately transforming the creation operators, which is done via \cite{fabre_modes_2020}\begin{equation}
    \hat{a}^{\dagger}_n = \sum_{k=0}^{\infty}(\chi_n, \chi_k')\hat{b}^\dagger_k,
\end{equation} where $(\cdot,\cdot)$ is an inner product over the modal space. For the remainder of this work the we take the inner product to be given by equation (\ref{eq:ip}).

Let us now fix the $\chi'$-basis as being given by equation (\ref{eq:modes}), and calculate the derivative of $\hat{a}^{\dagger}_0$ with respect to $\alpha_j$ when the $\chi$-basis also matches equation (\ref{eq:modes}). We find \begin{subequations}
\begin{align}
    \frac{\partial\hat{a}^{\dagger}_0}{\partial\alpha_j} &= \frac{\partial}{\partial\alpha_j}\sum_{k=0}^{\infty}(\omega_n, \omega_k')\hat{a}^\dagger_k \\
    &= \sum_{k=0}^{\infty}\left(\frac{\partial\omega_n}{\partial\alpha_j}, \omega_k'\right)\hat{a}^\dagger_k \\
    &= \sum_{k=0}^{\infty}\left(\omega_j, \omega_k'\right)\hat{a}^\dagger_k \\
    &= \sum_{k=0}^{\infty}\delta_{jk}\hat{a}^\dagger_k \\
    &= \hat{a}^\dagger_j.
\end{align}\end{subequations} Moreover, since $[\hat{a}^{\dagger}_0, \hat{a}^{\dagger}_j] = 0$ we have \begin{equation}
    \frac{\partial}{\partial\alpha_j}\left(\hat{a}^{\dagger}_0\right)^n = n\left(\hat{a}^{\dagger}_0\right)^{n-1}\hat{a}^{\dagger}_j.
\end{equation} The latter will be useful for evaluating the derivative of the density matrix in Appendix \ref{app:dsnQFI}, and the derivative of the single-mode squeezed vacuum state in Appendix \ref{app:derivative-smsv}.

\subsection{Derivative of the Single-Mode Squeezed Vacuum with Respect to Modal Phase Coefficients}
\label{app:derivative-smsv}
A single-mode squeezed vacuum state has the number-state representation \cite{loudon2000quantum} \begin{align}
    \ket{0,\xi}_0 &= \hat{S}(\xi)\ket{0}_0 \\
                &= \frac{1}{\sqrt{\cosh(r)}}\sum_{n=0}^{\infty}\frac{\sqrt{(2n)!}}{2^nn!}\left[-e^{i\theta}\tanh(r)\right]^n\ket{2n}_0 \\
                &= \frac{1}{\sqrt{\cosh(r)}}\sum_{n=0}^{\infty}\frac{(-1)^n}{2^n n!}e^{in\theta}\tanh^n(r)(\hat{a}^{\dagger}_0)^{2n}\ket{0}_0.
\end{align} Using the results of Appendix \ref{app:derivative_creation} we can calculate the derivative of a single-mode squeezed vacuum with respect to the k$^\text{th}$ modal coefficient as \begin{widetext}\begin{subequations}
\begin{align}
    \frac{\partial}{\partial\alpha_k}\ket{0, \xi}_0 &= \frac{1}{\sqrt{\cosh(r)}}\sum_{n=0}^{\infty}\frac{(-1)^n}{2^n n!}e^{in\theta}\tanh^n(r)\frac{\partial(\hat{a}^{\dagger}_0)^{2n}}{\partial\alpha_k}\ket{0}_0 \\
    &= \frac{1}{\sqrt{\cosh(r)}}\sum_{n=1}^{\infty}\frac{(-1)^n}{2^n (n-1)!}e^{in\theta}\tanh^n(r)(\hat{a}^{\dagger}_0)^{2n-1}\ket{0}_0\ket{1}_k \\
    &= \frac{1}{\sqrt{\cosh(r)}}\sum_{n=0}^{\infty}\frac{(-1)^{n+1}}{2^n n!}e^{(n+1)i\theta}\tanh^{n+1}(r)(\hat{a}^{\dagger}_0)^{2(n-1)}\ket{0}_0\ket{1}_k \\
    &= -e^{i\theta}\tanh(r)\hat{a}^{\dagger}_0\hat{S}(\xi)\ket{0}_0\ket{1}_k \\
    &= -e^{i\theta}\tanh(r)\hat{S}(\xi)\hat{S}^{\dagger}(\xi)\hat{a}^{\dagger}_0\hat{S}(\xi)\ket{0}_0\ket{1}_k \\
    &= -e^{i\theta}\sinh(r)\hat{S}(\xi)\ket{1}_0\ket{1}_k
\end{align} \end{subequations}\end{widetext} where the last line follows from use of equation (\ref{eq:bog-S_normal}).

\subsection{Derivative of the Displacement Operator with respect to Modal Phase Coefficients}
\label{app:derivative-coherent}
We recall that the operators $\hat{a}_k$ and $\hat{a}^{\dagger}_l$ commute for $k\neq l$. Hence \begin{equation}
    \frac{\partial}{\partial\alpha_k}\hat{D}(\mu) = (\mu\hat{a}^{\dagger}_k-\mu^*\hat{a}_k) \hat{D}(\mu).
\end{equation} When acting on states of the form $\ket{n}_0\ket{0}_k$, this operator yields \begin{equation}
    \left[\frac{\partial}{\partial\alpha_k}\hat{D}(\mu)\right]\ket{0} = \mu\left[\hat{D}(\mu)\ket{n}_0\right]\ket{1}_k.
\end{equation}

\section{Derivation of the QFI for a Statistical Mixture of Displaced Squeezed Number States}
\label{app:dsnQFI}
\subsection{Derivative of the Density Operator}
\label{eq:dsn_d_dOp}
Let us use the Bogoliubov transforms for the operators $\hat{D}(\mu)$ and $\hat{S}(\xi)$ to rewrite the density operator as \begin{widetext}
\begin{align}
    \hat{\rho} &= \sum_{n=0}^{\infty}\epsilon_n\hat{D}(\mu)\hat{S}(\xi)\ket{n}_0\bra{n}_0\hat{S}^{\dagger}(\xi)\hat{D}^{\dagger}(\mu) \\
            &= \sum_{n=0}^{\infty}\frac{\epsilon_n}{n!}\hat{D}(\mu)\hat{S}(\xi)(\hat{a}^{\dagger}_0)^n\vacp\hat{a}_0^n\hat{S}^{\dagger}(\xi)\hat{D}^{\dagger}(\mu) \\
            &= \sum_{n=0}^{\infty}\frac{\epsilon_n}{n!} \hat{D}(\mu)\left[\hat{S}(\xi)\hat{a}^{\dagger}_0\hat{S}^{\dagger}(\xi)\right]^n\hat{S}(\xi)\vacp\hat{S}^{\dagger}(\xi)\left[\hat{S}(\xi)\hat{a}_0\hat{S}^{\dagger}(\xi)\right]^n\hat{D}^{\dagger}(\mu) \\
            &= \sum_{n=0}^{\infty}\frac{\epsilon_n}{n!} \left[\hat{D}(\mu)\hat{S}(\xi)\hat{a}^{\dagger}_0\hat{S}^{\dagger}(\xi)\hat{D}^{\dagger}(\mu)\right]^n\hat{D}(\mu)\hat{S}(\xi)\vacp\hat{S}^{\dagger}(\xi)\hat{D}^{\dagger}(\mu)\left[\hat{D}(\mu)\hat{S}(\xi)\hat{a}_0\hat{S}^{\dagger}(\xi)\hat{D}^{\dagger}(\mu)\right]^n.
\end{align} \end{widetext} The operator \begin{equation}
    \hat{g}^{\dagger}_0 = \hat{D}(\mu)\hat{S}(\xi)\hat{a}^{\dagger}_0\hat{S}^{\dagger}(\xi)\hat{D}^{\dagger}(\mu)
\end{equation} may be found by double application of the Bogoliubov transformation \cite{moller_displaced_1996} {\allowdisplaybreaks\begin{align}
    \hat{g}^{\dagger}_0 &= \hat{D}(\mu)[\hat{a}^{\dagger}_0\cosh(r)+\hat{a}_0e^{i\theta}\sinh(r)]\hat{D}^{\dagger}(\mu) \\
    &= \hat{D}(\mu)\hat{a}^{\dagger}_0\hat{D}^{\dagger}(\mu)\cosh(r)+\hat{D}(\mu)\hat{a}_0\hat{D}^{\dagger}(\mu)e^{i\theta}\sinh(r) \\
    &= (\hat{a}^{\dagger}_0 - \mu^*)\cosh(r)+(\hat{a}_0-\mu)e^{i\theta}\sinh(r).
\end{align}} The density operator can now be written in the notationally simpler form \begin{equation}
    \label{eq:dOp-g_notation}
    \hat{\rho} = \sum_{n=0}^{\infty}\frac{\epsilon_n}{n!} (\hat{g}^{\dagger}_0)^n\hat{D}(\mu)\hat{S}(\xi)\vacp\hat{S}^{\dagger}(\xi)\hat{D}^{\dagger}(\mu)g_0^n.
\end{equation}

Let us now differentiate equation (\ref{eq:dOp-g_notation}) with respect to the coefficient $\alpha_k$: \begin{widetext}\begin{align}
    \frac{\partial\hat{\rho}}{\partial\alpha_k} &= \sum_{n=0}^{\infty}\frac{\epsilon_n}{n!} \left[n\cosh(r)(\hat{g}^{\dagger}_0)^{n-1}\hat{D}(\mu)\hat{S}(\xi)\ket{0}_0\ket{1}_k\bra{0}_k\bra{0}_0\hat{S}^{\dagger}(\xi)\hat{D}^{\dagger}(\mu)g_0^n \right. \\
    &+ \mu(\hat{g}^{\dagger}_0)^{n}\hat{D}(\mu)\hat{S}(\xi)\ket{0}_0\ket{1}_k\bra{0}_k\bra{0}_0\hat{S}^{\dagger}(\xi)\hat{D}^{\dagger}(\mu)g_0^n \nonumber \\
    &-e^{i\theta}\sinh(r) (\hat{g}^{\dagger}_0)^{n}\hat{D}(\mu)\hat{S}(\xi)\ket{1}_0\ket{1}_k\bra{0}_k\bra{0}_0\hat{S}^{\dagger}(\xi)\hat{D}^{\dagger}(\mu)g_0^n \nonumber \\
    &-e^{-i\theta}\sinh(r) (\hat{g}^{\dagger}_0)^{n}\hat{D}(\mu)\hat{S}(\xi)\ket{0}_0\ket{0}_k\bra{1}_k\bra{1}_0\hat{S}^{\dagger}(\xi)\hat{D}^{\dagger}(\mu)g_0^n\nonumber \\
    &+ \mu^*(\hat{g}^{\dagger}_0)^{n}\hat{D}(\mu)\hat{S}(\xi)\ket{0}_0\ket{0}_k\bra{1}_k\bra{0}_0\hat{S}^{\dagger}(\xi)\hat{D}^{\dagger}(\mu)g_0^n \nonumber \\
    &+ \left.n\cosh(r)(\hat{g}^{\dagger}_0)^{n}\hat{D}(\mu)\hat{S}(\xi)\ket{0}_0\ket{0}_k\bra{1}_k\bra{0}_0\hat{S}^{\dagger}(\xi)\hat{D}^{\dagger}(\mu)g_0^{n-1} \right] \\
    &= \sum_{n=0}^{\infty}\epsilon_n \sqrt{n}\cosh(r)\left[\ket{n-1}_0'\ket{1}_k\bra{0}_k\bra{n}_0' + \ket{n}_0'\ket{0}_0\bra{1}_k\bra{n-1}_0'\right] \\
    &+\sum_{n=0}^{\infty}\epsilon_n \left[\mu\ket{n}_0'\ket{1}_k\bra{0}_k\bra{n}_0' + \mu^*\ket{n}_0'\ket{0}_k\bra{1}_k\bra{n}_0'\right] \nonumber \\
    &-\sum_{n=0}^{\infty}\epsilon_n \sqrt{n+1}\sinh(r)\left[e^{i\theta}\ket{n+1}_0'\ket{1}_k\bra{0}_k\bra{n}_0' + e^{-i\theta}\ket{n}_0'\ket{0}_k\bra{1}_k\bra{n+1}_0'\right].\nonumber
\end{align} \end{widetext} Notice that the only non-zero matrix elements of the derivative are those associated to ket/bra pairs of the form ($\{\ket{n}_0'\ket{1}_k\}_{n=0} ^{\infty}\}, \{\bra{n}_0'\bra{0}_k\}_{n=0}^{\infty})$ and ($\{\ket{n}_0'\ket{0}_k\}_{n=0}^{\infty}\}, \{\bra{n}_0'\bra{1}_k\}_{n=0}^{\infty}\}$). These elements are given by \begin{widetext}\begin{align}
    \label{eq:dsn_mat1}
    \bra{n}_0'\bra{1}_k\partial_k\hat{\rho}\ket{0}_k\ket{m}_0' & = \epsilon_m\left[\sqrt{m}\cosh(r)\delta_{n(m-1)} + \mu\delta_{nm}-e^{i\theta}\sinh(r)\delta_{n(m+1)}\right] \\
    \label{eq:dsn_mat2}
    \bra{m}_0'\bra{0}_k\partial_k\hat{\rho}\ket{1}_k\ket{n}_0' & = \epsilon_m\left[\sqrt{m}\cosh(r)\delta_{n(m-1)} + \mu^*\delta_{nm}-e^{-i\theta}\sinh(r)\delta_{n(m+1)}\right].
\end{align} \end{widetext}

\subsection{Symmetric Logarithmic Derivatives}
\label{app:dsn_sld}
Having both the eigenvectors of $\hat{\rho}$ and the matrix elements of $\partial_k\hat{\rho}$ in the eigenbasis, we can calculate the symmetric logarithmic derivatives from \cite{liu_quantum_2020} \begin{equation}
    \label{app:eq:sld-def}
    \hat{\mathfrak{L}}_\nu = 2\sum_{\lambda_k+\lambda_l \neq0}\frac{\bra{\lambda_k}\partial_{\nu}\hat{\rho}\ket{\lambda_l}}{\lambda_k + \lambda_l}\ket{\lambda_k}\bra{\lambda_l},
\end{equation} where $\lambda_k$ is an eigenvalue of $\hat{\rho}$ and $\ket{\lambda_k}$ the associated eigenstate. Defining \begin{equation}
    \tilde{\epsilon}_n = \begin{cases}
        0, \epsilon_n = 0 \\
        1, \epsilon\neq0
    \end{cases},
\end{equation} substitution into equation (\ref{app:eq:sld-def}) gives \begin{widetext} \begin{align}
    \hat{\mathfrak{L}}_\nu &= 2\sum_{n=1}\sum_{m=0}\tilde{\epsilon}_m\sqrt{m}\cosh(r)\left( 
    \ket{n-1}'_0\ket{1}_{\nu}\bra{0}_{\nu}\bra{m}'_0 + \ket{m}'_0\ket{0}_{\nu}\bra{1}_{\nu}\bra{n-1}'_0
    \right)\delta_{nm} \nonumber\\
    & +2\sum_{n=0}\sum_{m=0}\tilde{\epsilon}_m\left(  
    \mu\ket{n}'_0\ket{1}_{\nu}\bra{0}_{\nu}\bra{m}'_0 + \mu^*\ket{m}'_0\ket{0}_{\nu}\bra{1}_{\nu}\bra{n}'_0
    \right)\delta_{nm} \nonumber\\
    & -2\sum_{n=0}\sum_{m=1}\tilde{\epsilon}_{m-1}\sqrt{m}\sinh(r)\left(  
    e^{i\theta}\ket{n}'_0\ket{1}_{\nu}\bra{0}_{\nu}\bra{m-1}'_0 + e^{-i\theta}\ket{m-1}'_0\ket{0}_{\nu}\bra{1}_{\nu}\bra{n}'_0
    \right)\delta_{nm} \\
    %%%%%%%%
    &= 2\sum_{n=1}\tilde{\epsilon}_n\sqrt{n}\cosh(r)\left( 
    \ket{n-1}'_0\ket{1}_{\nu}\bra{0}_{\nu}\bra{n}'_0 + \ket{n}'_0\ket{0}_{\nu}\bra{1}_{\nu}\bra{n-1}'_0
    \right) \nonumber\\
    & +2\sum_{n=0}\tilde{\epsilon}_n\left(  
    \mu\ket{n}'_0\ket{1}_{\nu}\bra{0}_{\nu}\bra{n}'_0 + \mu^*\ket{n}'_0\ket{0}_{\nu}\bra{1}_{\nu}\bra{n}'_0
    \right) \nonumber\\
    & -2\sum_{m=1}\tilde{\epsilon}_{m-1}\sqrt{m}\sinh(r)\left(  
    e^{i\theta}\ket{m}'_0\ket{1}_{\nu}\bra{0}_{\nu}\bra{m-1}'_0 + e^{-i\theta}\ket{m-1}'_0\ket{0}_{\nu}\bra{1}_{\nu}\bra{m}'_0
    \right) \\
    %%%%%
    &= 2\sum_{n=1}\tilde{\epsilon}_n\sqrt{n}\cosh(r)\left( 
    \ket{n-1}'_0\ket{1}_{\nu}\bra{0}_{\nu}\bra{n}'_0 + \ket{n}'_0\ket{0}_{\nu}\bra{1}_{\nu}\bra{n-1}'_0
    \right) \nonumber\\
    & +2\sum_{n=0}\tilde{\epsilon}_n\left(  
    \mu\ket{n}'_0\ket{1}_{\nu}\bra{0}_{\nu}\bra{n}'_0 + \mu^*\ket{n}'_0\ket{0}_{\nu}\bra{1}_{\nu}\bra{n}'_0
    \right) \nonumber\\
    & -2\sum_{n=0}\tilde{\epsilon}_{n}\sqrt{n+1}\sinh(r)\left(  
    e^{i\theta}\ket{n+1}'_0\ket{1}_{\nu}\bra{0}_{\nu}\bra{n}'_0 + e^{-i\theta}\ket{n}'_0\ket{0}_{\nu}\bra{1}_{\nu}\bra{n+1}'_0
    \right) \\
    %%%%%
    &= 2\sum_{n=0}\tilde{\epsilon}_n\sqrt{n}\cosh(r)\left( 
    \ket{n-1}'_0\ket{1}_{\nu}\bra{0}_{\nu}\bra{n}'_0 + \ket{n}'_0\ket{0}_{\nu}\bra{1}_{\nu}\bra{n-1}'_0
    \right) \nonumber\\
    & +2\sum_{n=0}\tilde{\epsilon}_n\left(  
    \mu\ket{n}'_0\ket{1}_{\nu}\bra{0}_{\nu}\bra{n}'_0 + \mu^*\ket{n}'_0\ket{0}_{\nu}\bra{1}_{\nu}\bra{n}'_0
    \right) \nonumber\\
    & -2\sum_{n=0}\tilde{\epsilon}_{n}\sqrt{n+1}\sinh(r)\left(  
    e^{i\theta}\ket{n+1}'_0\ket{1}_{\nu}\bra{0}_{\nu}\bra{n}'_0 + e^{-i\theta}\ket{n}'_0\ket{0}_{\nu}\bra{1}_{\nu}\bra{n+1}'_0
    \right),
\end{align}
\end{widetext} where in moving to the last equality we have introduced the convention $\ket{-1}'_0 = \ket{0}'_0$.

For calculating the QFIM, and for verifying weak commutativity, we now evaluate $\hat{\rho}\hat{\mathfrak{L}}_{\nu}$. We start by realizing that the product of $\hat{\rho}$ with the first term in each of the summands of $\hat{\mathfrak{L}}_{\nu}$ vanishes. Thus, we can restrict our attention to the second terms, finding \begin{widetext}
    \begin{align}
        \hat{\rho}\hat{\mathfrak{L}}_{\nu} &= 2\left(\sum_{k=0}^{\infty}\epsilon_k\ket{k}'_0\bra{k}'_0\right)\sum_{n=0}^{\infty}\tilde{\epsilon}_n\left\{  
        \sqrt{n}\cosh(r)\ket{n}'_0\ket{0}_{\nu}\bra{1}_{\nu}\bra{n-1}'_0 + \mu^*\ket{n}'_0\ket{0}_{\nu}\bra{1}_{\nu}\bra{n}'_0 \right. \nonumber\\
        &\left. -\sqrt{n+1}\sinh(r)e^{-i\theta}\ket{n}'_0\ket{0}_{\nu}\bra{1}_{\nu}\bra{n+1}'_0
        \right\} \\
        %%%%%%
        &= 2\sum_{n=0}^{\infty}\sum_{k=0}^{\infty}\tilde{\epsilon}_n\epsilon_k\left\{  
        \sqrt{n}\cosh(r)\ket{k}'_0\ket{0}_{\nu}\bra{1}_{\nu}\bra{n-1}'_0 + \mu^*\ket{k}'_0\ket{0}_{\nu}\bra{1}_{\nu}\bra{n}'_0 \right. \nonumber\\
        &\left. -\sqrt{n+1}\sinh(r)e^{-i\theta}\ket{k}'_0\ket{0}_{\nu}\bra{1}_{\nu}\bra{n+1}'_0
        \right\}\delta_{nk} \\
        %%%%
        &= 2\sum_{n=0}^{\infty}\epsilon_n\left\{  
        \sqrt{n}\cosh(r)\ket{n}'_0\ket{0}_{\nu}\bra{1}_{\nu}\bra{n-1}'_0 + \mu^*\ket{n}'_0\ket{0}_{\nu}\bra{1}_{\nu}\bra{n}'_0 \right. \nonumber\\
        &\left.- \sqrt{n+1}\sinh(r)e^{-i\theta}\ket{n}'_0\ket{0}_{\nu}\bra{1}_{\nu}\bra{n+1}'_0
        \right\} 
    \end{align}
\end{widetext}

Weak commutativity follows immediately from this result. Right multiplying by $\hat{\mathfrak{L}}_{\nu'}$ ($\nu\neq\nu'$) will always give zero, since $\bra{1}_{\nu}\cdot\ket{1}_{\nu'} = \delta_{\nu\nu'}.$ Hence, \begin{align}
    \text{Tr}\left(\hat{\rho}\left[\hat{\mathfrak{L}}_{\nu}, \hat{\mathfrak{L}}_{\nu'}\right]\right) &= \text{Tr}\left(\hat{\rho}\hat{\mathfrak{L}}_{\nu}\hat{\mathfrak{L}}_{\nu'}\right) - \text{Tr}\left(\hat{\rho}\hat{\mathfrak{L}}_{\nu'}\hat{\mathfrak{L}}_{\nu}\right)\\
    &= \text{Tr}\left(\hat{\rho}\hat{\mathfrak{L}}_{\nu}^2\right)(\delta_{\nu\nu'} - \delta_{\nu'\nu}) \\
    &= 0.
\end{align}

\subsection{QFIM}
\label{app:dsn_qfim}
We can finally calculate the QFIM. Given the results of the prior section, we need only the products $\hat{\rho}\hat{\mathfrak{L}}_{\nu}^2$. Multiplying out yields \begin{widetext}
    \begin{align}
    \hat{\rho}\hat{\mathfrak{L}}_{\nu}^2 &= 2\sum_{n=0}^{\infty}\epsilon_n\ket{n}'_0\ket{0}_{\nu}\left\{\sqrt{n}\cosh(r)\bra{1}_{\nu}\bra{n-1}'_0 + \mu^*\bra{1}_{\nu}\bra{n}'_0 - \sqrt{n+1}e^{-i\theta}\sinh(r)\bra{1}_{\nu}\bra{n+1}'_0\right\} \\
    &\times2\sum_{m=0}^{\infty}\left\{\sqrt{m}\cosh(r)\ket{m-1}_0\ket{1}_{\nu} + \mu\ket{m}'_0\ket{1}_{\nu} - \sqrt{m+1}\sinh(r)e^{i\theta}\ket{m+1}'_0\ket{1}_{\nu}\right\}\bra{0}_{\nu}\bra{m}'_0 \nonumber\\
    &= 4\sum_{n=0}^{\infty}\sum_{m=0}^{\infty}\epsilon_n\tilde{\epsilon}_m\left\{
    \sqrt{nm}\cosh^2(r)\delta_{nm} + \mu\sqrt{n}\cosh(r)\delta_{m(n-1)} - \frac{\sqrt{n(m+1)}}{2}\sinh(2r)e^{i\theta}\delta_{(n-1)(m+1)}
    \right. \nonumber\\
    &\left.  
    +\mu^*\sqrt{m}\cosh(r)\delta_{n(m-1)} + |\mu|^2\delta_{nm} - \mu^*\sqrt{m+1}e^{i\theta}\sinh(r)\delta_{n(m+1)} - \frac{\sqrt{m(n+1)}}{2}e^{-i\theta}\sinh(2r)\delta_{(n+1)(m-1)}
    \right. \nonumber\\
    &\left.  
    -\mu^*\sqrt{n+1}e^{i\theta}\sinh(r)\delta_{(n+1)m} + \sqrt{(n+1)(m+1)}\sinh^2(r)\delta_{nm}
    \right\} \ket{n}'_0\bra{m}'_0.
    \end{align}
\end{widetext} We will restrict our attention to terms of the form $\ket{n}'_0\bra{n}'_0$, as only these will survive tracing. The QFIM is therefore
\begin{widetext}
\begin{align}
    Q_{\nu\nu'} &= \text{ReTr}\left(\hat{\rho}\hat{\mathfrak{L}}_{\nu}\hat{\mathfrak{L}}_{\nu'}\right) \\
    &= \text{ReTr}\left(\hat{\rho}\hat{\mathfrak{L}}_{\nu}^2\right)\delta_{\nu\nu'}\\
            &= 4\sum_{n=0}^{\infty}\epsilon_n\left[n\cosh^2(r) + |\mu^2| + (n+1)\sinh^2(r)\right]\delta_{\nu\nu'} \\
            &= 4\left[\langle n\rangle\cosh^2(r) + |\mu^2| + (\langle n\rangle + 1)\sinh^2(r)\right]\delta_{\nu\nu'} \\
            &= 4\left[\langle n\rangle\cosh(2r) + |\mu^2| + \sinh^2(r)\right]\delta_{\nu\nu'},
\end{align}\end{widetext} where \begin{equation}
    \langle n\rangle = \sum_{n=0}^{\infty}n\epsilon_n.
\end{equation}

%%%%%%%%%%%%%%%%
\section{Quantum Fisher Information for Statistical Mixtures of Squeezed and Displaced Number States}
\label{app:sdn-qfi}
We define a mixed squeezed and displaced number states as one of the form \begin{equation}
    \label{eq:sdn}
    \rho' = \hat{S}(\xi)\hat{D}(\tau)\left(\sum_{n=0}^{\infty}\epsilon_n'\ket{n}_0\bra{n}_0\right)\hat{D}(\tau)\hat{S}(\xi),
\end{equation} where the $\epsilon_n'$ are real positive constants that sum to unity, and $\tau=\tau_r+i\tau_i$. By use of the relation \cite{moller_displaced_1996} \begin{equation}
    \hat{D}(\tau)\hat{S}(\xi)=\hat{S}(\xi)\hat{D}(\tau'),
\end{equation} where \begin{equation}
    \label{eq:dsn-sdn}
    \tau' = \tau\cosh(r)+\tau^*e^{i\theta}\sinh(r)
\end{equation} we can cast the density operator as that of an equivalent mixed DSN state. Hence, the QFIM for mixtures of squeezed and displaced number states may be calculated may be calculated using equation (\ref{eq:qfi}), but now with $\mu$ given by \begin{align}
    \mu &= e^{-2r}\left\{[\cosh(r)+\sinh(r)\cos(\theta)]\tau_r + \sinh(r)\sin(\theta)\tau_i\right\} \nonumber \\
    &+ie^{-2r}\left\{[\cosh(r)+\sinh(r)\cos(\theta)]\tau_i -\sinh(r)\sin(\theta)\tau_r\right\}.
\end{align} This expression is easily derived by inverting equation (\ref{eq:dsn-sdn}).

%%%%%%%%%%%%
\section{Constructing the Optimal Measurement for Thermal States}
\label{app:thermal-povm}
We first construct the optimal POVM for a pure state input: $\ket{n}_0$. Let the kets $\{\ket{n_k}_0\}_{k=1}^{\infty}$ represent the derivative of $\ket{n}_0$ with respect to the $k^{\text{th}}$ modal coefficient, and $U\in\mathbb{R}^{\infty\times\infty}$ be an orthogonal matrix expressed in the orthonormal basis $\{\ket{n}_0, \partial_1\ket{n}_0, \partial_2\ket{n}_0, \dots\}$. The optimal POVM ($\Pi_{n}$) is \cite{pezze_optimal_2017} \begin{equation}
    \Pi_n = \left\{\ket{\Upsilon_{nk}}\bra{\Upsilon_{nk}}\right\}_{k=0}^\infty \cup \left\{\mathbb{I} - \sum_{k=0}^{\infty}\ket{\Upsilon_{nk}}\bra{\Upsilon_{nk}}\right\},
\end{equation} where the kets $\ket{\Upsilon_{nk}}$ are defined as \begin{equation}
    \label{app:eq:optimal-kets-thermal}
    \ket{\Upsilon_{nk}} = U_{0k}\ket{n}_0 + \sum_{l=1}^{\infty}U_{lk}\partial_l\ket{n}_0,\,\,U_{0k}\neq0.
\end{equation}

One now notices that $\braket{\Upsilon_{nk}}{\Upsilon_{ml}} = \delta_{nm}\delta_{kl}$. This implies that the subspaces spanned by $\{\ket{\Upsilon_{nk}}\}_{k=0}^{\infty}$ and $\{\ket{\Upsilon_{mk}}\}_{k=0}^{\infty}$ are pairwise orthogonal. We can therefore construct the optimal POVM for the general mixed state by simply combining the projectors over these orthogonal spaces as \begin{align}
    \label{app:eq:optimal-thermal-povm}
    \Pi &= \{\vacp\} \cup\left\{\bigcup_{n=1}^{\infty} \left\{\ket{\Upsilon_{nk}}\bra{\Upsilon_{nk}}\right\}_{k=0}^\infty \right\} \nonumber\\
    &\cup 
    \left\{\mathbb{I} - \vacp - \sum_{n=1}\sum_{k=0}^{\infty}\ket{\Upsilon_{nk}}\bra{\Upsilon_{nk}}\right\}.
\end{align} 

Optimality of this POVM follows immediately. Realizing that the first and last operators defined in equation (\ref{app:eq:optimal-thermal-povm}) contribute no information, we calculate the $(i,j)^{\text{th}}$ element of the CFIM ($F$) as
\begin{subequations}\allowdisplaybreaks
    \begin{align}
    F_{ij} &= \sum_{n=1}^{\infty}\sum_{m=0}^{\infty}\frac{\text{Tr}\left(\partial_i\hat{\rho}\ket{\Upsilon_{nm}}\bra{\Upsilon_{nm}}\right)\text{Tr}\left(\partial_j\hat{\rho}\ket{\Upsilon_{nm}}\bra{\Upsilon_{nm}}\right)}{\text{Tr}\left(\hat{\rho}\ket{\Upsilon_{nm}}\bra{\Upsilon_{nm}}\right)} \\
    &= 4\sum_{n=1}^{\infty}\sum_{m=0}^{\infty}\sum_{p=1}^{N\infty}\sum_{q=1}^{\infty}\frac{\sqrt{pq}\epsilon_p\epsilon_q}{\epsilon_nU_{0n}^2} \nonumber \\
    &\times\text{Re}[\bra{p}_0\bra{1}_i\cdot\ket{\Upsilon_{nm}}\bra{\Upsilon_{nm}}_i\cdot\ket{p}_0\ket{0}_i] \nonumber\\
    & \times\text{Re}[\bra{p}_0\bra{1}_j\cdot\ket{\Upsilon_{nm}}\bra{\Upsilon_{nm}}_i\cdot\ket{p}_0\ket{0}_j] \\
    &= 4\sum_{n=1}^{\infty}\sum_{m=0}^{\infty}\sum_{p=1}^{N\infty}\sum_{q=1}^{\infty} \sqrt{pq}\epsilon_p \epsilon_q U_{im}U_{jm}\delta_{np}\delta_{nq} / \epsilon_n \\
    &= 4\left(\sum_{n=1}^{\infty}n\epsilon_n\right)\left(\sum_{m=0}^{\infty}U_{im}U_{jm}\right) \\
    &= 4\langle n\rangle\delta_{ij}
\end{align}
\end{subequations} which proves $F=F^Q$. Of particular note is the reliance of this construction only on the orthogonality of subspaces. This same process is thus applicable to any density operator for which the optimal kets correspoding to the constituent pure states (eq. \ref{app:eq:optimal-kets-thermal}) are orthonormal.

\section{Quantum-Optimal Measurements for Coherent States}
Let $\{\ket{\Upsilon_k}\bra{\Upsilon_k}\}_{k=0}^{\infty}$ denote the optimal POVM for measuring the wavefront of an coherent state with displacement $\mu$ $(\neq0)$. These kets may be constructed according the the sum \begin{equation}
    \ket{\Upsilon_k} = \sum_{l=0}^{\infty}U_{lk}\begin{cases}
        \ket{\psi_0}, & l = 0\\
        \frac{1}{\mu}\ket{\psi_l}, & l \neq 0
    \end{cases},
\end{equation} where $\{\ket{\psi_i}\}_{i=1}^\infty$ are the derivatives of $\ket{\psi_0}$ with respect to $\alpha_i$, the $U_{lk}$ are elements of an orthogonal matrix, and $U_{0k}\neq0$. Division by $\mu$ is necessary to normalize the state. 

A straightforward evaluation of the CFI now proves optimality: \begin{subequations}
    \begin{align}
        F_{ij} &= 4\sum_{n=0}^{N}\frac{\text{Re}[\braket{\psi_i(\mu)}{\Upsilon_n}\braket{\Upsilon_n}{1}]}{|\braket{\Upsilon_n}{\psi_0}|^2}\\
        &\hspace{2cm}\times\text{Re}[\braket{1}{\Upsilon_n}\braket{\Upsilon_n}{\psi_0}]\nonumber\\
        &= 4|\mu|^2\sum_{n=0}^{N}U_{in}U_{jn} \\
        &= 4|\mu|^2\delta_{ij}.
    \end{align}
\end{subequations}

\section{CFIM for Detuned Coherent State Measurements}
\label{app:cfim-coherent-nuisance}
Suppose that we are taking measurements with a coherent state probe whose true parameter is $\mu$, but due to imprecise characterization have prepared an optimal POVM for the coherent state with parameter $\mu'$ $(\neq0)$. The classical Fisher information, $F$, for applying the detuned optimal measurement to the input is \begin{subequations}\allowdisplaybreaks
    \begin{align}
        F_{ij} &= 4\sum_{n=0}^{N}\frac{\text{Re}[\braket{\psi_i}{\Upsilon_n}\braket{\Upsilon_n}{\psi_0}]\text{Re}[\braket{\psi_j}{\Upsilon_n}\braket{\Upsilon_n}{\psi_0}]}{|\braket{\Upsilon_n}{\psi_0}|^2} \\
         &= 4|\mu|^2\sum_{n=0}^{N}U_{in}U_{jn}|\braket{\mu, 0}{\mu', 0}|^2 \\
         &= 4|\mu|^2\cdot|\braket{\mu, 0}{\mu', 0}|^2\delta_{ij}.
    \end{align}
\end{subequations} The inner products are given by \cite{moller_displaced_1996} \begin{equation}
    \braket{\mu, 0}{\mu', 0} = e^{-\left[|\mu|^2+|\mu'|^2 - 2\text{Re}(\mu^*\mu')\right]/2}.
\end{equation} Hence the CFIM is diagonal with elements given by \begin{equation}
    F_{ij} = 4|\mu|^2e^{-|\mu-\mu'|^2}\delta_{ij}.
\end{equation}

\section{Quantum Optimal Measurements for Squeezed Vacuum States and the Effects of Imperfect State Characterization}
\label{app:cfim-squeezed-nuisance}
Suppose that we are taking measurements with a squeezed vacuum probe whose squeezing parameter is \begin{align}
    \xi &= re^{i\zeta},
\end{align} but due to imperfect characterization have prepared an optimal POVM for the squeezed state with parameter \begin{align}
    \xi' &= r'e^{i\zeta'} \\
        &\neq0\nonumber.
\end{align} The (thought to be) optimal POVM $\{\ket{\Upsilon_k}\bra{\Upsilon_k}\}_k$ is constructed using \begin{equation}
    \label{app:eq:inoptimal-squeeze}
    \ket{\Upsilon_k} = \sum_{l=0}^{N}U_{lk}\begin{cases}
        \ket{\psi_0}, & l = 0\\
        -\frac{e^{-i\theta}}{\sinh(r')}\ket{\psi_l}, & l \neq 0
    \end{cases},
\end{equation} where $\{\ket{\psi_i}\}_{i=1}^\infty$ are the derivatives of the input state, $\ket{\psi_0}$, with respect to $\alpha_i$, and the $U_{lk}$ are elements of an orthogonal matrix.

The classical Fisher information, $F$, of applying the measurement \ref{app:eq:inoptimal-squeeze} is \begin{subequations}\allowdisplaybreaks
    \begin{align}
        F_{ij} &= 4\sum_{n=0}^{N}\frac{\text{Re}[\braket{\psi_i}{\Upsilon_n}\braket{\Upsilon_n}{\psi_0}]\text{Re}[\braket{\psi_j}{\Upsilon_n}\braket{\Upsilon_n}{\psi_0}]}{|\braket{\Upsilon_n}{\psi_0}|^2} \\
         &= 4\sinh^2(r)\sum_{n=0}^{N}U_{in}U_{jn}|\braket{0, \xi}{0, \xi'}|^2 \\
         &= 4\sinh^2(r)\cdot|\braket{0, \xi, 1}{0, \xi', 1}|^2\delta_{ij},
    \end{align}
\end{subequations} where the $1$ inside the kets indicates that the squeezing operator is applied to the single photon state. The inner products are now evaluated as \cite{moller_displaced_1996} \begin{widetext}
    \begin{equation}
    \braket{0, \xi,1}{0, \xi',1} = \left|\cosh(r)\cosh(r') - e^{-i(\zeta-\zeta')}\sinh(r)\sinh(r')\right|^{-3/2}.
\end{equation}
\end{widetext} The CFIM is thus diagonal with elements given by 
\begin{widetext}
    \begin{align}
    F_{ij} &= \frac{4\sinh^2(r)\delta_{ij}}{\left|\cosh(r)\cosh(r') - e^{-i(\zeta-\zeta')}\sinh(r)\sinh(r')\right|^3}.
\end{align}
\end{widetext} Notice that when $\xi=\xi'$ the CFIM is equal to the QFIM.


\begin{thebibliography}{10}

    \bibitem{guyon2018extreme}
    Olivier Guyon.
    \newblock Extreme adaptive optics.
    \newblock {\em Annual Review of Astronomy and Astrophysics}, 56:315--355, 2018.
    
    \bibitem{briers1999optical}
    J~David Briers.
    \newblock Optical testing: a review and tutorial for optical engineers.
    \newblock {\em Optics and lasers in engineering}, 32(2):111--138, 1999.
    
    \bibitem{booth2014adaptive}
    Martin~J Booth.
    \newblock Adaptive optical microscopy: the ongoing quest for a perfect image.
    \newblock {\em Light: Science \& Applications}, 3(4):e165--e165, 2014.
    
    \bibitem{millane1990phase}
    Rick~P Millane.
    \newblock Phase retrieval in crystallography and optics.
    \newblock {\em JOSA A}, 7(3):394--411, 1990.
    
    \bibitem{caves_quantum-mechanical_nodate}
    Carlton~M. Caves.
    \newblock Quantum-mechanical noise in an interferometer.
    \newblock {\em Phys. Rev. D}, 23:1693--1708, Apr 1981.
    
    \bibitem{pinel_ultimate_2012}
    Olivier Pinel, Julien Fade, Daniel Braun, Pu~Jian, Nicolas Treps, and Claude Fabre.
    \newblock Ultimate sensitivity of precision measurements with intense {Gaussian} quantum light: {A} multimodal approach.
    \newblock {\em Physical Review A}, 85(1):010101, January 2012.
    
    \bibitem{jarzyna_quantum_2012}
    Marcin Jarzyna and Rafał Demkowicz-Dobrzański.
    \newblock Quantum interferometry with and without an external phase reference.
    \newblock {\em Physical Review A}, 85(1):011801, January 2012.
    
    \bibitem{demkowicz-dobrzanski_elusive_2012}
    Rafał Demkowicz-Dobrzański, Jan Kołodyński, and Mădălin Guţă.
    \newblock The elusive {Heisenberg} limit in quantum-enhanced metrology.
    \newblock {\em Nature Communications}, 3(1):1063, September 2012.
    
    \bibitem{humphreys_quantum_2013}
    Peter~C. Humphreys, Marco Barbieri, Animesh Datta, and Ian~A. Walmsley.
    \newblock Quantum {Enhanced} {Multiple} {Phase} {Estimation}.
    \newblock {\em Physical Review Letters}, 111(7):070403, August 2013.
    
    \bibitem{sparaciari_bounds_2015}
    Carlo Sparaciari, Stefano Olivares, and Matteo G.~A. Paris.
    \newblock Bounds to precision for quantum interferometry with {Gaussian} states and operations.
    \newblock {\em Journal of the Optical Society of America B}, 32(7):1354, July 2015.
    
    \bibitem{nichols_multiparameter_2018}
    Rosanna Nichols, Pietro Liuzzo-Scorpo, Paul~A. Knott, and Gerardo Adesso.
    \newblock Multiparameter {Gaussian} quantum metrology.
    \newblock {\em Physical Review A}, 98(1):012114, July 2018.
    
    \bibitem{haffert2023}
    Sebastiaan~Y. Haffert, Jared~R. Males, and Olivier Guyon.
    \newblock Reaching the fundamental sensitivity limit of wavefront sensing on arbitrary apertures with the phase induced amplitude apodized zernike wavefront sensor (piaa-zwfs), 2023.
    
    \bibitem{villegas_optimal_2024}
    Arturo Villegas, M.~H.~M. Passos, Silvania~F. Pereira, and Juan~P. Torres.
    \newblock Optimal parameter estimation of shaped phase objects.
    \newblock {\em Physical Review A}, 109(3):032617, March 2024.
    
    \bibitem{pezze_optimal_2017}
    Luca Pezzè, Mario~A. Ciampini, Nicolò Spagnolo, Peter~C. Humphreys, Animesh Datta, Ian~A. Walmsley, Marco Barbieri, Fabio Sciarrino, and Augusto Smerzi.
    \newblock Optimal {Measurements} for {Simultaneous} {Quantum} {Estimation} of {Multiple} {Phases}.
    \newblock {\em Physical Review Letters}, 119(13):130504, September 2017.
    
    \bibitem{weedbrook_gaussian_2012}
    Christian Weedbrook, Stefano Pirandola, Raúl García-Patrón, Nicolas~J. Cerf, Timothy~C. Ralph, Jeffrey~H. Shapiro, and Seth Lloyd.
    \newblock Gaussian quantum information.
    \newblock {\em Reviews of Modern Physics}, 84(2):621--669, May 2012.
    
    \bibitem{braunstein_quantum_2005}
    Samuel~L Braunstein and Peter van Loock.
    \newblock Quantum information with continuous variables.
    \newblock {\em Quantum information with continuous variables}, 77(2), 2005.
    
    \bibitem{liu_quantum_2020}
    Jing Liu, Haidong Yuan, Xiao-Ming Lu, and Xiaoguang Wang.
    \newblock Quantum {Fisher} information matrix and multiparameter estimation.
    \newblock {\em Journal of Physics A: Mathematical and Theoretical}, 53(2):023001, January 2020.
    
    \bibitem{ragy_compatibility_2016}
    Sammy Ragy, Marcin Jarzyna, and Rafał Demkowicz-Dobrzański.
    \newblock Compatibility in multiparameter quantum metrology.
    \newblock {\em Physical Review A}, 94(5):052108, November 2016.
    
    \bibitem{demkowicz-dobrzanski_multi-parameter_2020}
    Rafał Demkowicz-Dobrzański, Wojciech Górecki, and Mădălin Guţă.
    \newblock Multi-parameter estimation beyond quantum {Fisher} information.
    \newblock {\em Journal of Physics A: Mathematical and Theoretical}, 53(36):363001, September 2020.
    
    \bibitem{cochran-fi-utility}
    W.~G. Cochran.
    \newblock Experiments for nonlinear functions.
    \newblock {\em Journal of the American Statistical Association}, 68(344):771--781, 1973.
    
    \bibitem{nagaoka_quantum_2004}
    H~Nagaoka, M~Hayashi, R~D Gill, and S~Massar.
    \newblock Quantum {Cram}´er-{Rao} {Bound} in {Mixed} {States} {Model}.
    \newblock {\em Asymptotic Theory Of Quantum Statistical Inference}, 2004.
    
    \bibitem{okamoto_experimental_2012}
    Ryo Okamoto, Minako Iefuji, Satoshi Oyama, Koichi Yamagata, Hiroshi Imai, Akio Fujiwara, and Shigeki Takeuchi.
    \newblock Experimental {Demonstration} of {Adaptive} {Quantum} {State} {Estimation}.
    \newblock {\em Physical Review Letters}, 109(13):130404, September 2012.
    
    \bibitem{fujiwara_strong_2011}
    Akio Fujiwara.
    \newblock Strong consistency and asymptotic efficiency for adaptive quantum estimation problems.
    \newblock {\em Journal of Physics A: Mathematical and Theoretical}, 44(7):079501, February 2011.
    
    \bibitem{kimizu_adaptive_2024}
    Masataka Kimizu, Fuyuhiko Tanaka, and Akio Fujiwara.
    \newblock Adaptive quantum state estimation for two optical point sources.
    \newblock {\em Physical Review A}, 109(3):032434, March 2024.
    
    \bibitem{rodriguez-garcia_adaptive_2023}
    Marco~A. Rodriguez-Garcia and Francisco~E. Becerra.
    \newblock Adaptive {Phase} {Estimation} with {Squeezed} {Vacuum} {States} near the {Quantum} {Limit}.
    \newblock In {\em Optica {Quantum} 2.0 {Conference} and {Exhibition}}, page QTh2A.21, Denver, Colorado, 2023. Optica Publishing Group.
    
    \bibitem{ozer_reconfigurable_2022}
    Itay Ozer, Michael~R. Grace, and Saikat Guha.
    \newblock Reconfigurable {Spatial}-{Mode} {Sorter} for {Super}-{Resolution} {Imaging}.
    \newblock In {\em Conference on {Lasers} and {Electro}-{Optics}}, page JTh3A.28, San Jose, California, 2022. Optica Publishing Group.
    
    \bibitem{tsang_quantum_2016}
    Mankei Tsang, Ranjith Nair, and Xiao-Ming Lu.
    \newblock Quantum {Theory} of {Superresolution} for {Two} {Incoherent} {Optical} {Point} {Sources}.
    \newblock {\em Physical Review X}, 6(3):031033, August 2016.
    
    \bibitem{glauber_coherent_1963}
    Roy~J. Glauber.
    \newblock Coherent and {Incoherent} {States} of the {Radiation} {Field}.
    \newblock {\em Physical Review}, 131(6):2766--2788, September 1963.
    
    \bibitem{stoler_equivalence_1970}
    David Stoler.
    \newblock Equivalence {Classes} of {Minimum} {Uncertainty} {Packets}.
    \newblock {\em Physical Review D}, 1(12):3217--3219, June 1970.
    
    \bibitem{loudon2000quantum}
    Rodney Loudon.
    \newblock {\em The quantum theory of light}.
    \newblock OUP Oxford, 2000.
    
    \bibitem{fabre_modes_2020}
    C.~Fabre and N.~Treps.
    \newblock Modes and states in quantum optics.
    \newblock {\em Reviews of Modern Physics}, 92(3):035005, September 2020.
    
    \bibitem{moller_displaced_1996}
    K.~B. Mo/ller, T.~G. Jo/rgensen, and J.~P. Dahl.
    \newblock Displaced squeezed number states: {Position} space representation, inner product, and some applications.
    \newblock {\em Physical Review A}, 54(6):5378--5385, December 1996.
    
    \end{thebibliography}
\end{document}